
\def\a{\alpha}

\def\vp{\varphi}

\def\L{\Lambda}

\def\O{\Omega}

\font\tenbb=msym10
\font\sevenbb=msym7
\font\fivebb=msym5
\newfam\bbfam
\textfont\bbfam=\tenbb \scriptfont\bbfam=\sevenbb
\scriptscriptfont\bbfam=\fivebb

\def\Nc{{\cal N}}

\def\Pc{{\cal P}}

\def\Vc{{\cal V}}

\def\part{\partial}

\def\ts{\times}

\def\and{\mathop{\rm and}\nolimits}

\catcode`\@=11
\def\displaylinesno #1{\displ@y\halign{
\hbox to\displaywidth{$\@lign\hfil\displaystyle##\hfil$}&
\llap{$##$}\crcr#1\crcr}}

\def\ldisplaylinesno #1{\displ@y\halign{
\hbox to\displaywidth{$\@lign\hfil\displaystyle##\hfil$}&
\kern-\displaywidth\rlap{$##$}
\tabskip\displaywidth\crcr#1\crcr}}
\catcode`\@=12

\def\buildrel#1\over#2{\mathrel{
\mathop{\kern 0pt#2}\limits^{#1}}}

\def\build#1_#2^#3{\mathrel{
\mathop{\kern 0pt#1}\limits_{#2}^{#3}}}

\def\hfl#1#2{\smash{\mathop{\hbox to 6mm{\rightarrowfill}}
\limits^{\scriptstyle#1}_{\scriptstyle#2}}}

\def\hfll#1#2{\smash{\mathop{\hbox to 6mm{\leftarrowfill}}
\limits^{\scriptstyle#1}_{\scriptstyle#2}}}

\def\up#1{\raise 1ex\hbox{\sevenrm#1}}

\def\cqfd{\unskip\kern 6pt\penalty 500
\raise -2pt\hbox{\vrule\vbox to10pt{\hrule width 4pt
\vfill\hrule}\vrule}\par}

\def\signed#1 (#2){{\unskip\nobreak\hfil\penalty 50
\hskip 2em\null\nobreak\hfil\sl#1\/ \rm(#2)
\parfillskip=0pt\finalhyphendemerits=0\par}}

\newfam\bffam \textfont\bffam=\tenbf \scriptfont\bffam=\sevenbf
\scriptscriptfont\bffam=\fivebf
\def\bf{\fam\bffam\tenbf}

\def\cc#1{\hfill\kern .7em#1\kern .7em\hfill}

\catcode`\@=11
\def\system#1{\left\{\null\,\vcenter{\openup1\jot\m@th
\ialign{\strut\hfil$##$&$##$\hfil&&\enspace$##$\enspace&
\hfil$##$&$##$\hfil\crcr#1\crcr}}\right.}
\catcode`\@=12

\def\boxit#1#2{\setbox1=\hbox{\kern#1{#2}\kern#1}%
\dimen1=\ht1 \advance\dimen1 by #1 \dimen2=\dp1 \advance\dimen2 by #1
\setbox1=\hbox{\vrule height\dimen1 depth\dimen2\box1\vrule}%
\setbox1=\vbox{\hrule\box1\hrule}%
\advance\dimen1 by .4pt \ht1=\dimen1
\advance\dimen2 by .4pt \dp1=\dimen2 \box1\relax}

\font\twelvebf=cmbx12

\newcount\notenumber \notenumber =1
\def\note#1{\footnote{$^{\the\notenumber}$}{#1}%
\global\advance\notenumber by 1}

\magnification=1200
\overfullrule=0pt

\vglue 1cm

\centerline{\twelvebf Predictions from Quantum Cosmology}

\vglue 1cm

\centerline{Alexander Vilenkin\footnote{*)}{\sevenrm On leave from Tufts
University \smallskip e-mail address: AVILENKI@PEARL.TUFTS.EDU}}

\bigskip

\centerline{\sevenrm Institut des Hautes Etudes Scientifiques}

\centerline{\sevenrm 35, route de Chartres}

\centerline{\sevenrm 91440 Bures-sur-Yvette}

\centerline{\sevenrm FRANCE}

\vglue 4cm
\baselineskip 2 \normalbaselineskip

The world view suggested by quantum cosmology is that inflating
universes with all possible values of the fundamental constants are
spontaneously created out of nothing. I explore the consequences of the
assumption that we are a ``typical'' civilization living in this metauniverse.
The conclusions include inflation with an extremely flat potential and low
thermalization temperature, structure formation by topological defects, and an
appreciable cosmological constant.

\vfill\eject
\baselineskip 2 \normalbaselineskip

Why do the constants of Nature take the particular values
that they are observed to have in our universe? It certainly appears that the
constants have not been selected at random. Assuming that the particle masses
are bounded by the Planck mass $m_p$ and the coupling constants are $\build
<_{\sim}^{} 1$, one expects that a random selection would give all masses $\sim
m_p$ and all couplings $\sim 1$. The cosmological constant would then be $\L
\sim m_p^2$ and the corresponding vacuum energy $\rho_v \sim m_p^4$. In
contrast,
some of the particle masses are more than 20 orders of magnitude below $m_p$,
and the actual value of $\rho_v$ is $\build <_{\sim}^{} 10^{-120} \ m_p^4$. (I
use the system of units in which $\hbar =c =1$.)

\medskip

It has been argued [1] that the values of the constants
are, to a large degree, determined by anthropic considerations: these values
should be consistent with the existence of conscious observers who can wonder
about them. If one assumes that the production of heavy elements in stars
and
their dispersement in supernova explosions are essential for the evolution of
life, then one finds that this Anthropic Principle imposes surprisingly
stringent constraints on the electron, proton and neutron masses ($m_e$,
$m_{pr}$ and $m_n$), the $W$-boson mass $m_W$, and the fine structure constant
$e^2$. An anthropic bound on the cosmological constant can be obtained by
requiring that gravitationally bound systems are formed before the universe is
dominated by the vacuum energy [2].

\medskip

I should also mention the popular view that there exists a unique logically
consistent Theory of Everything and that all constants can in principle be
determined from that theory. The problem, however, is that the constants we
observe depend not only on the fundamental Lagrangian, but also on the vacuum
state, which is likely not to be unique. For example, in higher-dimensional
theories, like superstring theory, the constants in the four-dimensional world
depend on the way in which the extra dimensions are compactified. Moreover,
Coleman has argued [3] that all constants appearing in sub-Planckian physics
become totally undetermined due to Planck-scale wormholes connecting distant
regions of spacetime.

\medskip

Finally, it has been suggested that the explanation for the values of some
constants can be found in quantum cosmology. The wave function of the universe
gives a probability distribution for the constants which can be peaked at some
particular values [4]. Wormhole effects can also contribute an important factor
to the probability [5]. Smolin [6] has argued that new expanding regions of the
universe may be created as a result of gravitational collapse due to quantum
gravity effects. Assuming that the constants in these ``daughter'' regions
deviate slightly from their values in the ``mother'' region, he conjectured
that the observed values of the constants are determined by ``natural
selection'' for the values that maximize the production of black holes. Some
problems with this conjecture have been pointed out in Ref. [7].

\medskip

In this paper I would like to suggest a different approach to determining the
constants of Nature. This approach is not entirely new and has elements of
both anthropic principle and quantum cosmology. However, to my knowledge, it
has not been clearly formulated and its implications have not been
systematically explored. My approach is based on the picture of the universe
suggested by quantum cosmology and by the inflationary scenario. In this
picture, small closed universes spontaneously nucleate out of nothing, where
``nothing'' refers to the absence of not only matter, but also of space and
time [8]. All universes in this metauniverse are disconnected from one another
and generally have different values for some of the constants.  This variation
may be due to different compactification schemes, wormhole effects, etc.  We
shall not adopt any particular hypothesis and keep an open mind as to which
constants can be varied and what is the allowed range of their variation.

\medskip

After nucleation, the universes enter a state of
inflationary expansion.  It is driven by the potential energy of a scalar field
$\vp$, while the field slowly ``rolls down'' its potential $V(\vp)$. When $\vp$
reaches the steep portion of the potential at some $\vp \sim \vp_*$, its energy
 thermalizes, and inflation is
followed by the usual radiation-dominated expansion. The evolution of $\vp$
is influenced by quantum fluctuations, and as a result thermalization does not
occur simultaneously in different parts of the universe. In many models it can
be shown that at any time there are parts of the universe that are still
inflating [9,10].  Such eternally inflating universes have a beginning,
but have no end.

\medskip

We are one of the infinite number of civilizations living in thermalized
regions of the metauniverse. Although it may be tempting to believe that our
civilization is very special, the history of cosmology demonstrates that the
assumption of being average is often a fruitful hypothesis. I call this
assumption the Principle of Mediocrity. We shall see that, compared to the
traditional point of view, this principle gives a rather different perspective
on what is natural and what is not.

\medskip

The Principle of Mediocrity suggests that we think of ourselves as a
civilization randomly picked in the metauniverse.  Denoting by $\a_i$ the
constants of Nature that can vary from one universe to another, we can write
the corresponding
probability distribution as

$$d\Pc (\a) = Z^{-1} \ w_{\rm nucl} (\a) \ \Nc (\a) \ \build
\Pi_{i}^{}
\ d\a_i . \eqno (1)$$

\noindent Here, $w_{\rm nucl} (\a) \ \Pi \ d\a_i$ is the probability of
nucleation for an inflating universe with a given set of $\a_i$ in the
intervals
$d\a_i$, $\Nc (\a)$ is the average number of civilizations in such a universe
(in its entire history) [11], and $Z$ is a normalization factor.  We shall
interpret (1) as an {\it a priori} probability distribution for $\a_i$.

\medskip

The inflating part of the universe can be divided into a quantum region, $V
(\vp) > V_q$, where
the dynamics of the inflaton field $\vp$ is dominated by quantum fluctuations,
and slow-roll region, $V_* < V(\vp) < V_q$, where the evolution is essentially
deterministic. ($V_* = V(\vp_*)$ corresponds to the end of inflation).  The
values of $V_*$ and $V_q$ are model-dependent.
The inflationary expansion rate is given by
$H^2 =8\pi \ V(\vp) /3m_p^2$
and can be arbitrarily high if $V(\vp)$ is unbounded from above.
In order to extend the validity of the theory up to $V(\vp) \sim m_p^4$, one
can include one-loop matter corrtections to Einstein's action [12].  This may
be adequate if the number $N$ of matter fields  is large, $N>>1$.  Then
it can be
shown [13] that the resulting equations have no inflationary solutions for
$V(\vp) > V_{\rm max} \sim m_p^4 /N$.  The inflationary expansion rate is
therefore bounded from above [14], $H < H_{\rm max} \sim m_p /\sqrt {N}$.
Smaller values of $H_{\rm max}$ can be obtained in dilatonic and
higher-dimensional gravity models, or simply in models where $V(\vp)$ is
bounded from above (e.g., when $\vp$ is a cyclic variable and has a finite
range).  Here, we shall
assume that, for one reason or another, $H$
is bounded by some $H_{\rm max}$.  Eternal inflation is possible if $V_{\rm
max} > V_q$.

\medskip

Let us first assume that $V_{\rm max} < V_q$ in the whole range of variation of
$\a_i$, so that inflation is finite.  Very roughly, we can write

$$ \Nc (\a) \sim \Vc_* (\a) \nu_{\rm civ} (\a),  \eqno(2)$$

\noindent where $\Vc_*$ is the volume of the universe at the end of inflation
(that is, the 3-volume of the hypersurface $V(\vp) = V_*$), and $\nu_{\rm civ}$
is the average number of civilizations originating per unit volume $\Vc_*$.
The maximum of $\Vc_*$ is achieved by maximizing the highest value of the
potential $V_{\rm max}$ at which inflation starts and minimizing the slope of
$V(\vp)$ between $V_{\rm max}$ and $V_*$: the field $\vp$ takes longer to roll
down for a flatter potential.

\medskip

The cosmological literature abounds with remarks on the ``unnaturally'' flat
potentials required by inflationary scenarios. The slope of the potential is
severely constrained by the observed isotropy of the cosmic microwave
background.  With the Principle of Mediocrity, the situation is reversed:
flat is natural! Instead of asking why $V(\vp)$ is so flat, one should now ask
why it is not flatter.

\medskip

Let us now consider the role of other factors in (1).  The
calculation of $w_{\rm nucl} (\a)$ is a matter of some controversy. The
result
depends on one's choice of boundary conditions for the wave function of the
universe (see, e.g., [8,15]). Here we shall adopt the tunneling boundary
condition. Then the semiclassical nucleation probability is proportional to
$\exp (-\vert S\vert)$, where $S$ is the Euclidean action of the corresponding
instanton. In Einstein's gravity, $\vert S\vert =\pi m_p^2 /H_{\rm max}^2$,
and thus $w_{\rm nucl} (\a)$ favors large values of $V_{\rm max}$ and is not
sensitive to other parameters of the model [16].

\medskip

An important role in constraining the values of $\a_i$ is played by the ``human
factor'', $\nu_{\rm civ} (\a)$. We do not know what other forms of intelligent
life are
possible, but the Principle of Mediocrity favors the hypothesis that our form
is the most common in the metauniverse. The conditions required for life of our
type to exist (the low-energy physics based on the symmetry group $SU(3)\ts
SU(2) \ts U(1)$, the existence of stars and planets, supernova explosions) may
then fix, by order of magnitude, the values of $e^2$, $m_e$, $m_{\rm pr}$ and
$m_W$, as discussed in Ref. [1]. Anthropic considerations also impose a bound
on
the allowed flatness of the inflaton potential $V(\vp)$. If the potential is
too
flat, then the thermalization temperature after inflation is too low for
baryogenesis. The lowest temperature at which baryogenesis can still occur is
set by the electroweak scale, $T_{\rm min} \sim m_W$. Hence, if other
constraints do not interfere, we expect the
universe to thermalize at $T \sim m_W$.  Specific constraints on $V(\vp)$
depend
on the couplings of $\vp$ to other fields and can be easily obtained in
specific
models.

\medskip

Super-flat potentials required by the Principle of Mediocrity give rise to
density fluctuations which are many orders of magnitude below the strength
needed for structure formation. This means that the observed structures must
have been seeded by some other mechanism. The only alternative mechanism
suggested so far is based on topological defects: strings, global monopoles and
textures, which could be formed at a symmetry breaking phase transition [17].
The required symmetry breaking scale for the defects is $\eta \sim 10^{16}
{\rm Ge} V$. With ``natural'' (in the traditional sense) values of the
couplings,
the transition temperature is $T_c \sim \eta$, which is much higher than the
thermalization temperature $(T_{\rm th} \sim m_W)$, and no defects are formed
after inflation. It is possible for the phase transition to occur during
inflation, but the resulting defects are inflated away, unless the transition
is sufficiently close to the end of inflation. To arrange this requires some
fine-tuning of the constants.  However, the alternative is to have
thermalization at a much
higher temperature and to cut down on the amount of inflation. Since the
dependence of the volume factor $\Vc_*$ on the duration of
inflation is exponential, we
expect that the gain in the volume will more than compensate for the decrease
in ``$\a$-space''
due to the fine-tuning. We note also that in some supersymmetric
models the critical temperature of superheavy string formation can
``naturally'' be as low as $m_W$ [18].

\medskip

The symmetry breaking scale $\eta \sim 10^{16} {\rm Ge} V$ for the defects is
suggested by observations, but we have not explained why this particular scale
has been selected.  The value of $\eta$ determines the amplitude of density
fluctuations, which in turn determines the time when galaxies form, the
galactic density, and the rate of star formation in the galaxies.  Since these
parameters certainly affect the chances for civilizations to develop, it is
quite possible that $\eta$ is significantly constrained by the anthropic factor
$\nu_{\rm civ} (\a)$.

\medskip

If $\nu_{\rm civ}$ is indeed sharply peaked at some value of $\eta$ and thus
fixes the amplitude of density
fluctuations and the epoch of active galaxy formation, then an upper bound on
the cosmological constant can be obtained by requiring that it does not disrupt
galaxy formation until the end of that epoch. The growth of density
fluctuations in a flat universe with $\L >0$ effectively stops at a redshift
[19] $1+z\sim (1-\O_{\L} )^{-1/3}$, where $\O_{\L} =\rho_v /\rho_c$ and
$\rho_c$ is the critical density. Requiring that this happens at $z \build
<_{\sim}^{} 1$ gives $\O_{\L} \build <_{\sim}^{} 0.9$. The actual value of $\L$
is likely to be comparable to this upper bound.  Negative values of $\Lambda$
are bounded by requiring that our part of the universe does not recollapse
while stars are still shining and new civilizations are being formed.  This
gives a bound comparable to that for positive $\Lambda$ (by absolute value).

\medskip

Let us now turn to the case of eternal inflation, $V_{\rm max} > V_q$.  The
evolution of $\vp$ is then a stochastic process and can be described by a
distribution function $\rho (\vp , t)$ which satisfies a ``diffusion equation''
with appropriate boundary conditions at $V(\vp) = V_*$ and $V(\vp) = V_{\rm
max}$ [9,20-23].  In an eternally inflating universe, the volume $\Vc_*$ of the
hypersurfaces $V(\vp) = V_*$ is infinite and has to be regulated.  The simplest
way to do this is to cut it off at some time $t = \tau$ and consider the
asymptotic behavior as $\tau \to \infty$.  The time variable $t$ can be defined
as the proper time on the congruence of geodesics orthogonal to the initial
hypersurface at the ``moment of nucleation''.  Since geodesics tend to diverge
during inflation, this proper-time gauge should be well defined.
If $\rho$ is normalized to the total inflating volume, then
in the limit $t \to \infty$ we have [23] $\rho = F(\vp) \exp (dH_{\rm max} t)$,
 where $d(\a)$ can be interpreted [21]
as the fractal dimension of the region expanding at the highest rate
$H_{\rm max} (\a)$, $0 < d(\a) < 3$.  The asymptotic
form of $\Vc_*$ at large $\tau$ is then

$$ \Vc_* (\a, \tau) = {\tilde \Vc} (\a) \exp [d (\a) H_{\rm max} (\a)
\tau], \eqno(3)$$

\noindent and it is clear that in the limit $\tau \to \infty$ the distribution
(1) selects the values of $\a_i$ that maximize the product $d(\a)H_{\rm max}
(\a)$,

$$B(\a) \equiv d (\a) H_{\rm max} (\a) = {\rm max}.   \eqno(4)$$

Generically, a function attains its absolute maximum at a single point.  If
this is so for $B(\a)$, then Eq.(4) is sufficient to determine all constants of
Nature.  However, it is conceivable that the maximum of $B(\a)$ is degenerate,
so that (4) defines a surface in the space of
$\a_i$.  All values not on this surface have a vanishing probability, and the
probability distribution on the surface is
proportional to $w_{\rm nucl}(\a) {\tilde \Vc}(\a) \nu_{\rm civ} (\a)$.

The functions $B(\a)$ and ${\tilde \Vc}(\a)$ depend on the choice of the time
variable $t$ which is used to implement the cutoff.  For example, if instead of
the proper time we chose
${\tilde t} = \Vc_* (\a, t)$, then by construction the factor
$\Vc_*$ would be the same for all universes.
Here, we shall keep the proper time cutoff, which has a simple geometric and
physical meaning.  It favors the universes producing the largest number of
civilizations per unit time by the clocks of the co-moving observers.  The
cutoff-dependence of the results is nontheless an important issue and requires
further study [24].

\medskip

The fractal dimension $d(\a)$ increases as the potential $V(\vp)$ becomes
flatter [21, 23], and thus the condition (4) selects maximally flat potentials
with the highest value of $V_{\rm max}$.  In some models, maximization of
$B(\a)$ may drive the slope of $V(\vp)$ to zero; then no reasonable cosmology
is obtained.  The approach presented here can be meaningful only if the maximum
of $B(\a)$ corresponds to a non-trivial potential $V(\vp)$.  If we assume in
addition that this maximum is degenerate and defines a surface rather than a
single point, then
the probability maximum on that surface
is determined by the same considerations as in the case of finite inflation.
In
particular, the electroweak scale should not
exceed the thermalization temperature,
since otherwise no baryons would be formed.  A more detailed discussion of
$d\Pc (\a)$ in the case of eternal inflation will be given elsewhere.

\medskip

Let us now summarize the ``predictions'' of the Principle of Mediocrity.  The
preferred models have very flat inflaton potentials, thermalization and
baryogenesis at the electroweak scale, density fluctuations seeded by
topological defects and a non-negligible $\O_\L$ (as long as these features are
consistent with one another and with the constraint (4)
in the case of eternal inflation).

\bigskip

After this work was completed, I learned about the preprints by A.Albrecht [25]
and by J.Garcia-Bellido and A.Linde [26] which have some overlap with the
ideas presented here.
I am grateful to Brandon Carter and Alan Guth for discussions and to Thibault
Damour for his hospitality at I.H.E.S. where this work was completed. This
research was supported in part by the National Science Foundation.

\vfill\eject

\def\date{le\ {\the\day}\ \ifcase\month\or janvier\or
{f\'evrier}\or mars\or avril \or mai\or juin\or juillet\or
{ao\^ut}\or septembre\or octobre\or novembre\or {d\'ecembre}\fi
\ {\oldstyle\the\year}}

\def\a{\alpha}

\def\vp{\varphi}

\def\L{\Lambda}

\def\O{\Omega}

\font\tenbb=msym10

\font\sevenbb=msym7
\font\fivebb=msym5

\newfam\bbfam
\textfont\bbfam=\tenbb \scriptfont\bbfam=\sevenbb
\scriptscriptfont\bbfam=\fivebb

 \rm

\def\part{\partial}

\def\and{\mathop{\rm and}\nolimits}

\def\exp{\mathop{\rm exp}\nolimits}

\catcode`\@=11
\def\Eqalign#1{\null\,\vcenter{\openup\jot\m@th\ialign{
\strut\hfil$\displaystyle{##}$&$\displaystyle{{}##}$\hfil
&&\quad\strut\hfil$\displaystyle{##}$&$\displaystyle{{}##}$
\hfil\crcr#1\crcr}}\,} \catcode`\@=12

\catcode`\@=11
\def\displaylinesno #1{\displ@y\halign{
\hbox to\displaywidth{$\@lign\hfil\displaystyle##\hfil$}&
\llap{$##$}\crcr#1\crcr}}

\def\ldisplaylinesno #1{\displ@y\halign{
\hbox to\displaywidth{$\@lign\hfil\displaystyle##\hfil$}&
\kern-\displaywidth\rlap{$##$}
\tabskip\displaywidth\crcr#1\crcr}}
\catcode`\@=12

\def\buildrel#1\over#2{\mathrel{
\mathop{\kern 0pt#2}\limits^{#1}}}

\def\build#1_#2^#3{\mathrel{
\mathop{\kern 0pt#1}\limits_{#2}^{#3}}}

\def\hfl#1#2{\smash{\mathop{\hbox to 6mm{\rightarrowfill}}
\limits^{\scriptstyle#1}_{\scriptstyle#2}}}

\def\up#1{\raise 1ex\hbox{\sevenrm#1}}

\def\signed#1 (#2){{\unskip\nobreak\hfil\penalty 50
\hskip 2em\null\nobreak\hfil\sl#1\/ \rm(#2)
\parfillskip=0pt\finalhyphendemerits=0\par}}

\def\TeX{T\kern-.1667em\lower.5ex\hbox{E}\kern-.125em X}

\def\lsim{ {\raise -3mm \hbox{$<$} \atop \raise 2mm
\hbox{$\sim$}} }

\def\gsim{ {\raise -3mm \hbox{$>$} \atop \raise 2mm
\hbox{$\sim$}} }

\def\frac#1#2{\mathop{\scriptstyle#1\over\scriptstyle#2}\nolimits}

\def\fnote#1{\advance\noteno by 1\footnote{$^{\the\noteno}$}
{\eightpoint #1}}

\def\boxit#1#2{\setbox1=\hbox{\kern#1{#2}\kern#1}%
\dimen1=\ht1 \advance\dimen1 by #1 \dimen2=\dp1 \advance\dimen2 by
#1
\setbox1=\hbox{\vrule height\dimen1 depth\dimen2\box1\vrule}%
\setbox1=\vbox{\hrule\box1\hrule}%
\advance\dimen1 by .4pt \ht1=\dimen1
\advance\dimen2 by .4pt \dp1=\dimen2 \box1\relax}

\def\cube{
\raise 1 mm \hbox { $\boxit{3pt}{}$}
}

\def\cqfd{\unskip\kern 6pt\penalty 500
\raise -2pt\hbox{\vrule\vbox to10pt{\hrule width 4pt
\vfill\hrule}\vrule}\par}

\def\dstar {\displaystyle ({\raise- 2mm \hbox
{$*$} \atop \raise 2mm \hbox {$*$}})}

\def\ref #1#2{
\smallskip\parindent=2,0cm
\item{\hbox to\parindent{\enskip\lbrack{#1}\rbrack\hfill}}{#2} }

\def\choose#1#2{\mathop{\scriptstyle#1\choose\scriptstyle#2}\nolimits}

\def\adots{\mathinner{\mkern2mu\raise1pt\hbox{.}
\mkern3mu\raise4pt\hbox{.}\mkern1mu\raise7pt\hbox{.}}}

\def\pegal{\mathrel{\vbox{\hsize=9pt\hrule\kern1pt
\centerline {$\circ$}\kern.6pt\hrule}}}

\overfullrule=0mm

\def\gsim{ {\raise -3mm \hbox{$>$} \atop \raise 2mm
\hbox{$\sim$}} }

\centerline {\bf References}
\medskip

\ref{1.}{B.~Carter, in I.A.U. Symposium {\bf 63}, ed.~by M.S.~Longair (Reidel,
Dordrecht, 1974); Phil.~Trans.~R.~Soc.~Lond. {\bf A310}, 347 (1983);
B.J.~Carr and M.J.~Rees, Nature {\bf 278}, 605 (1979); J.D.~Barrow and
F.J.~Tipler, ``The Anthropic Cosmo-\break logical Principle" (Clarendon,
Oxford,
1986).  It should be noted that the Anthropic Principle, as originally
formulated by Carter, is more than a trivial consistency condition. It is the
requirement that anthropic constraints should be taken into account when
evaluating the plausibility of various hypotheses about the physical world.}

\ref {2.} {S.~Weinberg, Phys.~Rev.~Lett.~{\bf 59}, 2607 (1987)}

\ref {3.} {S.~Coleman, Nucl.~Phys.~{\bf B307}, 867 (1988).}

\ref {4.} {E.~Baum, Phys.~Lett.~{\bf B133}, 185 (1984); S.W.~Hawking,
Phys.~Lett.~{\bf B134}, 403 (1984).}

\ref {5.}{S.~Coleman, Nucl.~Phys.~{\bf B310}, 643 (1988). In this paper Coleman
obtained a probability distribution for  $\rho_v$ with an extremely
sharp peak at $\rho_v = 0$. However, his derivation was based on Euclidean
quantum gravity, which has serious problems. For a discussion of the problems,
see W.~Fischler et.~al., Nucl.~Phys.~{\bf B327}, 157 (1989).}

\ref {6.} {L.~Smolin, Class.~Quant.~Grav.~{\bf 9}, 173 (1992); Penn.~State
Preprint, un-\break published.}

\ref {7.} {T.~Rothman and G.F.R.~Ellis, Quart.~J.~Roy.~Astr.~Soc.~{\bf 34},
201 (1993).}

\ref {8.} {A.~Vilenkin, Phys.~Lett.~{\bf 117B}, 25 (1982), Phys.~Rev.~{\bf
D30},
509 (1984);\break J.~Hartle and S.W.~Hawking, Phys.~Rev.~{\bf D28}, 2960
(1983);
A.D.~Linde, Lett.~Nuovo Cim.~{\bf 39}, 401 (1984).}

\ref{9.} {A.~Vilenkin, Phys.~Rev.~{\bf D27}, 2848 (1983).}

\ref {10.} {A.D.~Linde, Phys.~Lett.~{\bf B175}, 395 (1986).}

\ref {11.} {We may wish to assign a weight to each civilization, depending on
its lifetime and/or on the number of individuals. This would not change the
conclusions of the present paper.}

\ref {12.} {See, e.g., A.A.~Starobinsky, Phys.~Lett.~{\bf 91B}, 99(1980).}

\ref {13.} {Including the vacuum contributions of matter fields to the
expectation value of $T_{\mu \nu}$ and assuming slow rollover conditions,
$\dot \vp^2 \ll V(\vp)$ and $\dot H \ll H^2$, the evolution equation for $H$
takes the form
$$H^2 = {8 \pi \over 3m^2_p} V(\vp) + {H^4 \over H_0^2} - {6 \over M^2} H^2
\dot H.$$
Here, $H_0 \sim m_p/\sqrt{N}$, $N$ is the number of matter fields with masses
$m \ll H$, and $M$ can be adjusted to any value by a finite renormalization of
the quadratic in curvature term in the gravitational Lagrangian. Physically
reasonable models are obtained for $H^2_0 > 0,\ M^2 > 0$ (for details see
Ref.~[12]). Classical inflationary solutions must have $\dot H < 0$. This
gives a quadratic inequality for $H^2$, which can be satisfied only if $V(\vp)
\le 3 H^2_0 m^2_p/32 \pi$. The expansion rate cannot exceed $H_0$. A detailed
discussion of this issue will be given elsewhere.}

\ref {14.} {Linde et. al. [23] have argued that the inflationary expansion
rate is bounded by $H_{\max} \sim m_p$, because at this rate quantum
fluctuations
in the energy-momentum tensor $T_\vp^{\mu \nu}$ of the inflaton field $\vp$
become comparable to $T_\vp ^{\mu \nu}$ itself, and the vacuum form of
$T_\vp^{\mu \nu} \propto g^{\mu \nu}$ is destroyed. I disagree with this
argument. At $H \sim m_p$, quantum fluctuations in $T^{\mu \nu}$ for all fields
with $m \ll m_p$ have comparable magnitude. The average total energy-momentum
tensor is $\langle T^{\mu \nu} \rangle \propto g_{\mu \nu}$, and its relative
fluctuation is $\sim N^{- 1/2}$, where $N$ is the number of fields with $m \ll
m_p$. Since $N \gsim 100$, the vacuum form of $T^{\mu \nu}$ holds with a good
accuracy.}

\ref {15.} {A.~Vilenkin, Phys.~Rev.~{\bf D37}, 888 (1988).}

\ref {16.} {$w_{\rm nucl}$ depends also on the initial value of $\vp$ and is
maximized at $V(\vp) = V_{\rm max}$.}

\ref {17.} {For a review of topological defects, see A.~Vilenkin and
E.P.S.~Shellard, ``Cosmic Strings and Other Topological Defects" (Cambridge
University Press, Cambridge, 1994).}

\ref {18.} {G.~Lazarides, C.~Panagiotakopoulos and Q.~Shafi,
Phys.~Rev.~Lett.~{\bf 56}, 432 (1987); Phys.~Lett.~{\bf 183B}, 289 (1987).}

\ref {19.} {S.M.~Carroll, W.H.~Press and E.L.~Turner,
Ann.~Rev.~Astron.~Astrophys.~{\bf 30}, 499 (1992).}

\ref {20.} {A.A.~Starobinsky, in ``Current Topics in Field Theory, Quantum
Gravity and Strings'', ed. by H.J.~de Vega and N.~Sanchez (Springer,
Heidelberg, 1986).}

\ref {21.} {M.~Aryal and A.~Vilenkin, Phys.~Lett.~{\bf B199}, 351 (1987)}

\ref {22.} {Y.~Nambu and M.~Sasaki, Phys.~Lett ~{\bf B219}, 240 (1989)}

\ref {23.} {A.D.~Linde and A.~Mezhlumian, Phys.~Lett.~{\bf B307}, 25 (1993);
A.D.~Linde, D.A.~Linde and A.~Mezhlumian, Phys.~Rev.~{\bf D49}, 1783 (1994).}

\ref {24.} {The dependence of $\rho (\vp, t)$ on the time parametrization has
been emphasized by Linde {\it et.al.} [23] and
by Garcia-Bellido {\it et.al.} [27]
who studied the probability distribution for the
inflaton and other fields in a single eternally inflating universe.}

\ref {25.} {A.Albrecht, Imperial College Report TP/93-94/56 (unpublished).}

\ref {26.} {J.~Garcia-Bellido and A.D.~Linde, Stanford University Report
SU-ITP-94-24 (unpublished)}

\ref {27.} {J.~Garcia-Bellido, A.D.~Linde and D.A.~Linde, Phys.~Rev.~{\bf D50},
730 (1994)}

\bye